\documentclass{llncs}

\usepackage[T1]{fontenc}
\usepackage[utf8]{inputenc}
\usepackage{todonotes}
\usepackage{url}
\usepackage{graphicx}
\usepackage{multirow}
\usepackage{booktabs}
\usepackage{cleveref}
\usepackage{microtype}

\begin{document}
\title{Reference String Extraction Using Line-Based Conditional Random Fields\thanks{Preprint}} 
\author{Martin K\"orner}

\titlerunning{Reference String Extraction}
\authorrunning{Martin K\"orner} 
\institute{Institute for Web Science and Technologies, University of Koblenz-Landau, Germany\\
  \email{mkoerner@uni-koblenz.de}
}
\maketitle             

\begin{abstract}
The extraction of individual reference strings from the reference section of scientific publications is an important step in the citation extraction pipeline.
Current approaches divide this task into two steps by first detecting the reference section areas and then grouping the text lines in such areas into reference strings.
We propose a classification model that considers every line in a publication as a potential part of a reference string.
By applying line-based conditional random fields rather than constructing the graphical model based on the individual words, dependencies and patterns that are typical in reference sections provide strong features while the overall complexity of the model is reduced.
\keywords{reference string extraction, conditional random fields}
\end{abstract}
	
\section{Introduction} 
A common approach for generating citation information is to extract them from research papers that are given in the PDF format.
In the following, we will consider the steps in the extraction process that result in individual reference strings.
 
There are several factors that result in the difficulty of the reference string extraction task.
One such factor is the high number of possible reference styles.
Further, there exist a large variety of layouts for publications including different section headings, headers, footers, and varying numbers of text columns.
Current solutions that perform reference string extraction have in common that they first identify the reference section and then, in a separate step, segment the reference section into individual reference strings.
Errors that are made during the classification of reference sections thereby directly impact the accuracy of the reference string extraction.
For example, if a paragraph that contains reference strings was not recognized as part of the reference section, those reference strings will not be considered in the following step.
To prevent this, our approach does not extract reference strings from an area that is first identified as the reference section.
Instead, this information is only used as one of many feature in a machine learning model that is applied on every textual line of a given publication. 
Other features are based on the text layout and the content of a given text line.
This approach was first presented at the EXCITE Workshop 2017 ``Challenges in Extracting and Managing References'' in Cologne, Germany \cite{ghavimi2017excite}.

\vspace{0.5cm}
\noindent
\textbf{Summary of Contributions}
\begin{itemize}
  \item Overview of related work in reference string extraction (\Cref{sec:related-work})
  \item New approach to reference string extraction using line-based supervised conditional random fields 
\end{itemize}

\section{Related Work}\label{sec:related-work}
There is a large body of research that addresses the extraction of bibliographic information from the reference section of research papers \cite{peng2004accurate,cortez2007flux,groza2012reference,lopez2009grobid,councill2008parscit,wu2014citeseerx,tkaczyk2015cermine}.
One group focuses on the reference string segmentation task by assuming the reference strings to be given \cite{peng2004accurate,cortez2007flux,groza2012reference} while another group also considers the reference string extraction from a given publication in the PDF or text format \cite{lopez2009grobid,councill2008parscit,wu2014citeseerx,tkaczyk2015cermine}.
Their reference string extraction approaches all have in common that their consist of two steps.

The first step identifies the text areas of the publication that contain the reference strings.
Given UTF-8 text files, Councill, Giles, and Kan \cite{councill2008parscit} as well as Wu et al. \cite{wu2014citeseerx} apply a set of regular expressions to locate the beginning and end of reference sections.
This relies on the heading of the reference section to be one of the predefined phrases such as ``References'', ``Bibliography'', or ``References and Notes''.
Further regular expressions are applied to detect the end of the reference section by searching for following sections such as figures, appendices, and acknowledgments \cite{councill2008parscit}.
Additionally, a detected reference section is ignored if it appears too early in the publication \cite{councill2008parscit}.
Tkaczyk et al. \cite{tkaczyk2015cermine} apply a layout analysis on publications given as PDF files which results in textual areas that are grouped into zones.
These zones are then classified as ``metadata'', ``body'', ``references'', or ``other'' using a trained Support Vector Machines (SVMs) model \cite{tkaczyk2015cermine}.
Lopez \cite{lopez2009grobid} trains a conditional random field (CRF) \cite{lafferty2001conditional} model that performs a segmentation of textual areas into zones similar to Tkaczyk et al. \cite{tkaczyk2015cermine}. 

In a second step, the lines in the identified areas are grouped into individual reference strings.
Councill, Giles, and Kan \cite{councill2008parscit} as well as Wu et al. \cite{wu2014citeseerx} apply regular expressions to detect possible markers of reference strings such as numbers or identifiers surrounded by brackets.
If such markers are found, the lines are grouped accordingly.
If no markers are found, the lines are grouped based on the line length, ending punctuation, and strings that appear to be author name lists\cite{councill2008parscit}.
Tkaczyk et al. \cite{tkaczyk2015cermine} use the k-means learning algorithm to perform a clustering into two groups: The first lines of research strings and all other lines.
The features for this clustering include layout information such as the distance to the previous line and textual information such as a line ending with a period\cite{tkaczyk2015cermine}.
As with the reference area detection, Lopez \cite{lopez2009grobid} learn a CRF model for this task.
This model uses an input format that is different from the one that is used for their first CRF model.
Tokens are split at whitespaces and for each token, a list of features is created.
Such features include layout information such as the font size and font weight as well as textual features such as the capitalization of the token and whether the token resembles a year, location, or name.

\section{Approach} 
A typical problem of existing reference string extraction approaches is a wrong classification of textual areas during the first step (see \Cref{sec:related-work}).
Areas which are not correctly identified as reference areas are not considered in the second step which potentially harms the recall of the extraction.
Areas which do not contain reference strings but are identified as reference areas on the other hand result in a loss of precision.
An approach that does not rely on the correct detection of reference areas thereby could be more robust.

Another key insight is that reference strings commonly start in a new line.
This assumption is used in the clustering algorithm of Tkaczyk et al. \cite{tkaczyk2015cermine} and potentially provides advantages over a word-based approach similar to the CRF model of Lopez \cite{lopez2009grobid}.
In addition to the reduced complexity of the model, a line-based approach can capture patterns that repeat every few lines more naturally than a word-based model which focuses on a more local context.

A line-based classification model that is applied on the whole publication combines the two insights.
For this, every line is classified as one of the following labels:
\begin{itemize}
  \item \texttt{B-REF}: The first line of a reference string.
  \item \texttt{I-REF}: A line of a reference string which is not the first line.
  \item \texttt{O-REF}: A line which appears inside a reference string but which is not part of it.
  \item \texttt{O}: Any other line.
\end{itemize}
The notation is based on the Beginning-Intermediate-Other (BIO) notation \cite{houngbo2012method}.
\Cref{tab:bio-example} shows a concrete example in which a page number appears in between the lines of a reference string which is split over two pages.
\begin{table}
\centering
\caption{Example of label assignments to a number of lines.}
\bgroup
\setlength{\tabcolsep}{10pt}
\begin{tabular}{l l}
 \toprule
 Label & Text Line\\
 \midrule
 \texttt{O} & grant numbers MA 3964/8-1 and STA 572/14-1.\\
 \texttt{O} & References\\
 \texttt{B-REF} & Tkaczyk, D., et al. (2015) Cermine: automatic extraction of\\
 \texttt{I-REF} & structured metadata from scientific literature. Interna-\\
 \texttt{O-REF} & 1252\\
 \texttt{I-REF} & tional Journal on Document Analysis and Recognition (IJDAR) 18(4) \\
 \texttt{B-REF} & Lafferty, J., McCallum, A., Pereira, F. (2001) Conditional random \\
 \bottomrule
\end{tabular}
\egroup
\label{tab:bio-example}
\end{table}

When having a line-based model, a variety of features can be computed for every line.
Features that are based on the layout leverage the information that is available when performing reference string extraction of publications in the PDF format. 
Examples for such layout features can be the space between the current line and the previous line, an existing indentation in comparison to the previous line, and information about the font size and weight.
In addition, the textual information of the line itself provides strong features.
They can focus on the appearance of a specific word or pattern such as a year, page range, or name in a given database.
It is also possible to provide features based on the start or end of the line.
Here it is interesting, for example, to assign features to lines that start with a number or end with a period.
Features that are based on the number of occurrences of punctuation marks or capitalized words can provide further indicators of a reference line.

One advantage of the conditional probability distribution that is modeled with CRFs is that there is no independence assumption needed between the features \cite{koller2009probabilistic}.
Another advantage is a possible leverage of typical patterns that appear in a reference section. 
To do so, a CRF model with a relatively high Markov order can be applied.
Because of the line-based approach and the resulting lower number of random variables in the model, the complexity of higher Markov orders is manageable when considering a supervised training with a relatively low amount of manually annotated documents.

\section{Conclusion and Future Work} 
We presented a simplified approach to the task of reference string extraction by applying one line-based classification model instead of using separate models for finding the reference section areas and grouping the lines into individual reference strings.
In addition to to reduced model complexity, this also simplifies the annotation process during the generation of training data.

One particularly interesting use case of the presented approach are publications that do not have a reference section at the end of the publication but contain the reference strings in the footnotes on the pages where the reference is made.
For example, a number of journals in the German social sciences such as ``Totalitarismus und Demokratie''\footnote{\url{http://www.hait.tu-dresden.de/td/home.asp}} and ``Südosteuropäische Hefte''\footnote{\url{http://suedosteuropaeische-hefte.org/}} use this way of referencing.

\section{Acknowledgments}
This work has been funded by Deutsche Forschungsgemeinschaft (DFG) as a part of the project “Extraction of Citations from PDF Documents (EXCITE)” under grant numbers MA 3964/8-1 and STA 572/14-1. 

\bibliographystyle{splncs}
\bibliography{bibdb.bib}

\end{document}